\documentclass[thmsa,10pt,a4paper,oneside,final,onecolumn,notitlepage,12pt]{article}
\usepackage{sw20lart}

\input tcilatex
\QQQ{Language}{
American English
}

\begin{document}

\title{Dirac Sea Contribution in Relativistic \\
Random Phase Approximation}
\author{S. S. Wu [1], H. X. Zhang [2] and Y. J. Yao [1] \\
1.Center for Theoretical Physics and School of Physics, \\
Jilin University, Changchun 130023, P. R. China\\
2. Institute of High Energy Physics, \\
The Chinese Academia of Sciences, Beijing 100049, P. R. China}
\date{}
\maketitle

\begin{abstract}
\baselineskip=0.6cm In the hadrodynamics (QHD) there are two methods to take
account of the contribution of negative-energy states in the relativistic
random phase approximation (RRPA). Dawson and Furnstahl made the ansatz that
the Dirac sea were empty, while according to the Dirac hole theory the sea
should be fully occupied. The two methods seem contradictory. Their close
relationship and compatibility are explored and in particular the question
of the ground-state (GS) instability resulting from Dawson-Furnstanhl's
ansatz is discussed.


\end{abstract}

\baselineskip=0.9cm

\newpage\ 

\section{Introduction}

In the framework of the conventional quantum hadrodynamics (CQHD), the
relativistic mean-field theory (RMFT) based on effective Lagrangians and the
no-sea approximation has achieved a remarkable success for the calculation
of nuclear ground states (referred to as MFGS i.e. mean-field GS) [1]. Since
the parameters are adjusted to fit the experimental data, one expects that a
large part of vacuum contributions and some other neglected effects are
already taken into account phenomenologically [1]. A formal justification of
the above conjecture has been given by Furnstahl, Pickarewicz and Serot [2]
on the basis of the effective field theory (EFT), where a more elaborate
effective Lagrangian with non-renormalizable interactions (modern QHD [3])
has been considered. For the calculation of excited states in the
relativistic random phase approximation (RRPA), Dawson and Furnstahl (D-F)
pointed out long ago [4] that in order to separate out the spurious $J^\pi
=1^{-}$ state and to preserve current conservation, it is of vital
importance to take the negative-energy (NE) states into account, because
only then will the Dirac single-particle basis become complete. They
suggested that based on the above MFGS one may assume the NE sea is empty
and thus besides the positive-energy (PE) particle-hole pairs one should
further consider pairs formed from a particle at one of the NE states and a
hole in the occupied PE states (referred to as $\alpha h$ pairs as in Ref.
[5]). Ma et al. [6] found that the D-F method can improve the calculation of
nuclear giant resonances significantly. Recently, Ring at al. [5] have made
a detailed study of the effects of the $\alpha h$ pairs and found that they
are indeed important. However, as is wellknown [7, 8], an empty NE sea
cannot avoid the problem of spontaneous radiation, i.e. of keeping the
valence nucleons from jumping into NE states through emitting mesons and
photons, if the corresponding interaction terms are included in the
Lagrangian. As numerous NE states are available, the MFGS, even though
successful in all the other aspects, will not be stable and will disappear
almost instantly. For this reason we would like to emphasize that it is
still important to follow Dirac's original proposal of a filled NE sea.
Evidently, a filled NE sea can equally fulfill all the above requirements
pointed out by D-F. Besides, if any effect due to the filled sea has been
encoded into the effective Lagrangian by fitting the parameters to
experimental data, in the calculation it should and can be neglected in
order to avoid redundancy (see section 2). Clearly, for such a purpose there
is no need to drive the sea particles out of the sea. Since the calculation
on the basis of a filled sea will be as simple as one with an empty sea if
the vacuum loops are neglected, it shows that Dirac's postulate is a
preferable choice for both CQHD and MQHD, because it cannot only prohibit
the valence nucleons from tumbling into the NE sea, but also take account of
the existence of antiparticles in a natural way. In addition, it also
provides a greater possibility to treat the vacuum contribution effectively,
namely if it is already included in the parameters of the effective
Lagrangian (EL), we may simply neglect it in the calculation; however, if a
part of it has been left out owing to a somewhat excessive truncation, we
may then either add additional terms and parameters to EL or calculate the
part as a contribution from the sea. It is known that the results of the
one-N (nucleon) -loop and virtual $N\overline{N\text{ }}$ pair calculations
based on structureless nucleons are questionable and earlier calculations
with CQHD met difficulties of large effects from loop integrals caused by
the vacuum contribution. It seems [9] that the above drawbacks can be
overcome by a careful study of vertex corrections and relevant higher-order
effects.

The relation between the effective D-F method and Dirac's hole theory will
be studied in some detail in section 2. In section 3 numerical results
calculated in RRPA are presented and discussed, where phenomenological
vertex corrections are also considered to make a preliminary study of the
effect due to the nucleon compositeness. Finally a summary is given in the
last section.

\section{Relation between D-F's effective method and Dirac's hole theory}

In order to expose the relation between D-F's effective method and Dirac's
hole theory more clearly and in simpler terms, we shall restrict our
discussion to symmetric nuclear matter and Walecka's $\sigma -\omega $
model. The Lagrangian is written as 
\begin{eqnarray}
L &=&-\overline{\psi }(\gamma _\mu \partial _\mu +M-g_s\sigma -ig_v\gamma
_\mu \omega _\mu )\psi  \nonumber \\
&&-\frac 12\partial _\mu \sigma \partial _\mu \sigma -\frac 12m_s^2\sigma
^2-\frac 14W_{\mu \nu }W_{\mu \nu }-\frac 12m_v^2\omega _\mu \omega _\mu 
\text{,}  \tag{1}
\end{eqnarray}
where $W_{\mu \nu }=\partial _\mu \omega _\nu -\partial _\nu \omega _\mu $
and $\psi $, $\sigma $, $\omega $ denote the nucleon, $\sigma $-meson and $%
\omega $-meson fields, respectively. As is wellknown [1], through fitting
the parameters to nuclear properties, the $\sigma -\omega $ model, though
simple, has achieved considerable success in the description of many bulk
and single-particle properties of nuclei. Thus we may assume that a part of
vacuum contributions and some higher-order effects have been included in the
Lagrangian $L$, though the inclusion cannot be so complete as what the more
elaborate effective Lagrangian may provide. We would like to show that if
the vacuum contribution from Dirac's filled NE sea is neglected, the Dirac
results reduce almost to what one obtains according to the D-F ansatz. In
order to explore the relationship between the two methods we may apply them
to calculate the same set of physical quantities and compare their results.
For this purpose we shall consider the correlation function 
\begin{equation}
C(A,B;x_1,x_2)=\left\langle T[A(x_1)B(x_2)]\right\rangle -\left\langle
A(x_1)\right\rangle \left\langle B(x_2)\right\rangle \text{,}  \tag{2}
\end{equation}
as a typical example. In Eq. (2) $\left\langle O\right\rangle =\left\langle
\Psi _0\left| O\right| \Psi _0\right\rangle ,$ $K(x)=\overline{\psi }%
(x)\Gamma _\kappa \psi (x)$ ($K=A$ or $B$), $\psi (x)$ is the nucleon field
operator, whereas operator $\Gamma _\kappa $ is field- and time-
independent. For instance, for the isovector multipole operator we have $%
\Gamma _\kappa =\gamma _4\tau _3r^\lambda Y_{\lambda \mu }(\theta ,\varphi )$
and for the bilinear Dirac current and scalar density $\Gamma _\kappa
=\gamma _\mu $ ($\kappa =\mu =1$, $2$, $3$, $4$) and $\Gamma _\kappa =1$ ($%
\kappa =5$), respectively, etc. For simplicity we shall further assume that $%
\Gamma _\kappa $ contains no differential operators. In the lowest order
approximation Eq. (2) has the form 
\begin{equation}
C^0(A,B;x_1,x_2)=-\mathbf{Tr}[\Gamma _A(\mathbf{x}_1)G^0(x_1-x_2)\Gamma _B(%
\mathbf{x}_2)G^0(x_2-x_1)]\text{,}  \tag{3}
\end{equation}
where $G^0(x)$ is the relativistic Hartree approximation to the nucleon
propagator 
\begin{equation}
G_{\alpha \beta }(x=x_1-x_2)=\left\langle T\mathbf{[}\psi _\alpha (x_1)%
\overline{\psi }_\beta (x_2)\mathbf{]}\right\rangle \text{,}  \tag{4}
\end{equation}
and $x\equiv x_\mu =(\mathbf{x},ix_0)$ with $x_0=t$. The Fourier transform
of Eq. (3) is given by 
\begin{eqnarray}
C^0(A,B;kk^{\prime }) &=&\diint d^4x_1d^4x_2e^{-ikx_1+ik^{\prime
}x_2}C^0(A,B;x_1,x_2)  \nonumber \\
\ &=&-2\pi \delta (k_0-k_0^{^{\prime }})\int \frac{d^4q}{(2\pi )^4}\frac{d^3p%
}{(2\pi )^3}\mathbf{Tr[}\Gamma _A(\mathbf{k}-\mathbf{k}^{\prime }-\mathbf{p})
\nonumber \\
&&\ \ \ \ \ \times G^0(\mathbf{k}^{\prime }+\mathbf{q}+\mathbf{p}%
,k_0^{\prime }+q_0)\Gamma _B(\mathbf{p})G^0(\mathbf{q},q_0)\mathbf{]}\text{.}
\tag{5a}
\end{eqnarray}
If $\Gamma _\kappa $ is further independent of $\mathbf{x}$ (indicated by $%
\kappa $ taking a small letter), we have $\Gamma _a(\mathbf{p})=(2\pi
)^3\delta ^3(p)\Gamma _a$ and Eq. (5a) reduces to 
\begin{eqnarray}
C^0(a,b;kk^{\prime }) &=&-(2\pi )^4\delta ^{(4)}(k-k^{\prime })\int \frac{%
d^4q}{(2\pi )^4}\mathbf{Tr}[\Gamma _aG^0(q+k)\Gamma _bG^0(q)]  \nonumber \\
\ &\equiv &(2\pi )^4\delta ^{(4)}(k-k^{\prime })C^0(a,b;k)\text{.}  \tag{5b}
\end{eqnarray}
One observes that except for the $\delta $-function and a constant factor,
Eq. (5b) is just the expression for the polarization tensor in the $\sigma
-\omega $ model, if $\Gamma _\kappa $ $(\kappa =1$ to $5)=\{\gamma _\mu ,1\}$%
. Since we are considering a nuclear matter whose GS is given by RMFT, $%
G^0(p)=G_F^0(p)+G_D^0(p)$ can be written as 
\begin{eqnarray}
G_F^0(p) &=&-(\gamma _\mu p_\mu +iM^{*})\left[ P\frac 1{p^2+M^{*2}}+i\pi
\delta (p^2+M^{*2})\right]  \tag{6a} \\
\ &=&(\gamma _\mu p_\mu +iM^{*})\frac 1{2|E_{\mathbf{p}}|}\{\frac{\theta (E_{%
\mathbf{p}})}{p_0-E_{\mathbf{p}}+i\epsilon }-\frac{\theta (-E_{\mathbf{p}})}{%
p_0-E_{\mathbf{p}}-i\epsilon }\}  \tag{6b}
\end{eqnarray}
\begin{equation}
G_D^0(p)=(\gamma _\mu p_\mu +iM^{*})\frac{i\pi }{E_{\mathbf{p}}}\theta
(p_0)\theta (k_F-|\mathbf{p}|)\delta (p_0-E_{\mathbf{p}})\text{,}  \tag{6c}
\end{equation}
where $E_{\mathbf{p}}=\pm [\mathbf{p}^2+M^{*2}]^{1/2}$, $M^{*}$ is the
effective mass, $k_F$ denotes the Fermi momentum, $F$ the Feynman part, $D$
the density-dependent part and $P$ the principle value, we may rewrite Eq.
(5) as 
\begin{equation}
C^0(A,B;kk^{\prime })=C_{FF}^0(A,B;kk^{\prime })+C_m^0(A,B;kk^{\prime
})+C_{DD}^0(A,B;kk^{\prime })\text{,}  \tag{7a}
\end{equation}
\begin{equation}
C_m^0(A,B;kk^{\prime })=C_{DF}^0(A,B;kk^{\prime })+C_{FD}^0(A,B;kk^{\prime })%
\text{.}  \tag{7b}
\end{equation}
On the right-hand side of Eq. (7) the subscript $D$ or $F$ indicates $G^0$
in Eq. (5) is $G_D^0$ or $G_F^0$ and the three parts will be referred to as $%
FF$-, $m$- and $DD$-part, respectively. It is wellknown that $\psi
(x)=e^{iHt}\psi (\mathbf{x},0)e^{-iHt}$ and $\psi (\mathbf{x},0)$ can be
expanded in the form 
\begin{equation}
\psi (\mathbf{x},0)=\sum_{r=1}^4\int \frac{d^3p}{(2\pi )^{3/2}}c(pr)u(pr)e^{i%
\mathbf{p\cdot x}}\text{,}  \tag{8}
\end{equation}
where we note $c(pr)=d^{+}(-p,r-2),$ $u(pr)=v(-p,r-2)$ if $r=3,$ $4$ and $%
v(pr)$ ($r=1,$ $2$) denotes a NE spinor with momentum $-\mathbf{p}$.
Substituting Eq. (8) into Eq. (4), we easily find [10] that the Fourier
transform of $G_{\alpha \beta }(x)$ can be represented in the form 
\begin{equation}
G_{\alpha \beta }(p)=\sum_{r,s=1}^4u_\alpha (pr)iG(pr,ps;p_0)\overline{u}%
_\beta (ps)\text{,}  \tag{9a}
\end{equation}
\begin{equation}
G(pr,ps;p_0)=\frac 1i\int_{-\infty }^{+\infty }dt\text{ }e^{ip_0t}\left%
\langle T\mathbf{[}c(pr,t_1)c^{+}(ps,t_2)\mathbf{]}\right\rangle \text{.} 
\tag{9b}
\end{equation}
Hereafter, we shall use a tilde to indicate the result obtained according to
the D-F method. D-F's ansatz [4] implies 
\begin{eqnarray}
\widetilde{G}^0(pr,ps;p_0) &=&\delta _{rs}\left[ \frac{\theta (E_{\mathbf{p}%
}-E_F)}{p_0-E_{\mathbf{p}}+i\epsilon }+\frac{\theta (-E_{\mathbf{p}})}{%
p_0-E_{\mathbf{p}}+i\epsilon }+\frac{\theta (E_F-E_{\mathbf{p}})\theta (E_{%
\mathbf{p}})}{p_0-E_{\mathbf{p}}-i\epsilon }\right]  \nonumber \\
&=&\delta _{rs}\widetilde{G}^0(p)\text{.}  \tag{10a}
\end{eqnarray}
Inserting Eq. (10a) in Eq. (9a), we obtain $\widetilde{G}^0(p)=\widetilde{G}%
_F^0(p)+\widetilde{G}_D^0(p),\widetilde{G}_D^0(p)=G_D^0(p)$ and 
\begin{equation}
\widetilde{G}_F^0(p)=(\gamma _\mu p_\mu +iM^{*})\frac 1{2|E_{\mathbf{p}%
}|}\left\{ \frac{\theta (E_{\mathbf{p}})}{p_0-E_{\mathbf{p}}+i\epsilon }-%
\frac{\theta (-E_{\mathbf{p}})}{p_0-E_{\mathbf{p}}+i\epsilon }\right\} \text{%
.}  \tag{10b}
\end{equation}
It is seen that the only difference between $G^0(p)$ and $\widetilde{G}^0(p)$
consists in their F part. In $\widetilde{G}_F^0(p)$ , just as $\widetilde{G}%
^0(p)$ in Eq. (10a), its NE pole is now shifted to the lower-half plane and
there are no other changes from $G_F^0(p)$ to $\widetilde{G}_F^0(p)$.
Defining $\Delta G^0(p)\equiv \widetilde{G}^0(p)-G^0(p)$, we have From Eqs.
(6) and (10) 
\begin{equation}
\Delta G^0(p)=\Delta G_F^0(p)=(\gamma _\mu p_\mu +iM^{*})\frac{i\pi }{|E_{%
\mathbf{p}}|}\delta (p_0-E_{\mathbf{p}})\theta (-E_{\mathbf{p}})\text{.} 
\tag{11}
\end{equation}
Since all the poles of $\widetilde{G}_F^0(p)$ are in the lower-half plane,
substituting Eq. (10) in Eq. (5), as pointed out in [4], we get $\widetilde{C%
}_{FF}^0(A,B;kk^{\prime })=0$ which also follows from the fact that here
there are no antiparticles. Besides, we find $\widetilde{C}%
_{DD}^0(A,B;kk^{\prime })=C_{DD}^0(A,B;kk^{\prime })$ and 
\begin{equation}
\widetilde{C}_m^0(A,B;kk^{\prime })=C_m^0(A,B;kk^{\prime })+\Delta C_m^0%
\text{,}  \tag{12a}
\end{equation}
\begin{equation}
\Delta C_m^0=C_{D,\Delta F}^0(A,B;kk^{\prime })+C_{\Delta
F,D}^0(A,B;kk^{\prime })\text{.}  \tag{12b}
\end{equation}
In Eq. (12b) the subscript $\Delta F$ means that we substitute $\Delta
G_F^0(p)$ for $G_F^0(p)$ in Eq. (7b). Comparing Eq. (12) with Eq. (7), we
have 
\begin{equation}
\Delta C^0\equiv \widetilde{C}^0-C^0=\Delta C_m^0-C_{FF}^0\text{,}  \tag{12c}
\end{equation}
i.e. the ansatz of empty NE sea is equivalent to the assertion that $%
C_{FF}^0 $ in $C^0$ is replaced by $\Delta C_m^0$. Thus, if the vacuum
contribution $C_{FF}^0$ can be neglected, the difference between the two
results consists solely of $\Delta C_m^0$. Note Eqs. (7) and (12) show why
the D-F method can give a better result than the no-sea approximation,
because the latter only takes account of $C_{DD}^0$ and a part of $C_m^0$
given by the first term in Eq. (6b) or (10b), while the former has further
correctly considered the full $C_m^0$ with even a correction $\Delta C_m^0$
(see below). In order to expose the physical implication of Eq. (12a) more
clearly, consider, for instance, Eq. (5b) with $\Gamma _a=\gamma _\eta $ and 
$\Gamma _b=\gamma _\lambda $. Substituting Eq. (6a) in Eq. (7b), and using a
prefix $\delta $ to denote the contribution of the second term of Eq. (6a),
we obtain 
\begin{equation}
\delta C_m^0(\eta ,\lambda ;k)=ReC_m^0(\eta ,\lambda ;k)=M_1(\eta ,\lambda
;k)+M_2(\eta ,\lambda ;k)\text{,}  \tag{13}
\end{equation}
\begin{eqnarray}
M_1(\eta ,\lambda ;k) &=&-\frac 1{8\pi ^2}\dint d^3q\frac{\theta (k_F-|%
\mathbf{q}|)}{E_{\mathbf{q}}}\{\frac{t_{\eta \lambda }(q+k)}{E_{\mathbf{q}+%
\mathbf{k}}}\delta (k_0+E_{\mathbf{q}}-E_{\mathbf{q}+\mathbf{k}})  \nonumber
\\
&&\ \ \ \ \ \ \ \ \ \ \ \ \ \ +\frac{t_{\eta \lambda }(q-k)}{E_{\mathbf{q}-%
\mathbf{k}}}\delta (k_0-E_{\mathbf{q}}+E_{\mathbf{q}-\mathbf{k}})\}\text{,} 
\tag{14a}
\end{eqnarray}
\begin{eqnarray}
M_2(\eta ,\lambda ;k) &=&-\frac 1{8\pi ^2}\dint d^3q\frac{\theta (k_F-|%
\mathbf{q}|)}{E_{\mathbf{q}}}\{\frac{t_{\eta \lambda }(q+k)}{E_{\mathbf{q}+%
\mathbf{k}}}\delta (k_0+E_{\mathbf{q}}+E_{\mathbf{q}+\mathbf{k}})  \nonumber
\\
&&\ \ \ \ \ \ \ \ \ \ \ \ \ \ \ +\frac{t_{\eta \lambda }(q-k)}{E_{\mathbf{q}-%
\mathbf{k}}}\delta (k_0-E_{\mathbf{q}}-E_{\mathbf{q}-\mathbf{k}})\}\text{,} 
\tag{14b}
\end{eqnarray}
where $Re$ means the real part and 
\begin{equation}
t_{\eta \lambda }(q\pm k)=[2q_\eta q_\lambda \pm q_\eta k_\lambda \pm
q_\lambda k_\eta \mp \delta _{\eta \lambda }(q\cdot k)]_{q_0=E_{\mathbf{q}}}%
\text{.}  \tag{15}
\end{equation}
From Eqs. (11-13), it is easily seen that 
\begin{equation}
\Delta C_m^0(\eta ,\lambda ;k)=-2M_2(\eta ,\lambda ;k)\text{,}  \tag{16}
\end{equation}
\begin{equation}
Re\widetilde{C}_m^0(\eta ,\lambda ;k)=ReC_m^0(\eta ,\lambda ;k)+\Delta
C_m^0(\eta ,\lambda ;k)=M_1(\eta ,\lambda ;k)-M_2(\eta ,\lambda ;k)\text{,} 
\tag{17a}
\end{equation}
\begin{equation}
Im\widetilde{C}_m^0(\eta ,\lambda ;k)=ImC_m^0(\eta ,\lambda ;k)\text{.} 
\tag{17b}
\end{equation}
According to Eqs. (16) and (17) one concludes that $\widetilde{C}_m^0$ and $%
C_m^0$ differ from each other only in the sign before $M_2$, because their
imaginary parts are the same. If we require $k_0>0$, it is seen from Eq.
(14b) that the first term in $M_2$ is zero and only the second term will
contribute. The $\delta $-function in Eq. (14b) shows $M_2$ becomes
effective only if $k_0>2M^{*}$, which is about $1.37GeV$ for $M^{*}\simeq
0.73M$, i.e. if the excitation energy is not too high, we have $\widetilde{C}%
_m^0\simeq C_m^0$. Since $M_2=0$ if $0<k_0<2M^{*}$, in the energy region
where for instance, giant multipole resonances are studied we even expect $%
\widetilde{C}_m^0=C_m^0$. Clearly the above derivation also applies to the
other components of $C^0(a,b;k)$ (see Eq. (5b)) as well as to Eq. (5a) and
similar results obtain. Sofar we have only considered the lowest order
approximation to Eq. (2). In order to gain some idea about the accumulative
effect, we have further made a RRPA calculation for $C(a,b;x_1,x_2)$ with $%
C^0$ given by Eq. (5b). The Dyson-Schwinger equation for the full
scalar-vector meson propagator in the $\sigma -\omega $ model can be written
in the form [11] 
\begin{equation}
\Delta _{ab}(k)=\Delta _{ab}^0(k)+\Delta _{ac}^0(k)\Pi _{cd}^0(k)\Delta
_{db}(k)\text{,}  \tag{18}
\end{equation}
where the polarization tensor $\Pi _{ab}^0(k)$ is related to $C^0(a,b;k)$ in
Eq. (5b) with $\Gamma _\kappa =\{\gamma _\mu ,1\}$ for $\kappa =a$ or $b$ as
follows: 
\begin{equation}
\Pi _{ab}^0(k)=\left( 
\begin{tabular}{ll}
$\Pi _{\mu \nu }^0$ & $\Pi _{\mu 5}^0$ \\ 
$\Pi _{5\nu }^0$ & $\Pi _{55}^0$%
\end{tabular}
\right) =\left( 
\begin{tabular}{ll}
$g_v^2C^0(\mu ,\nu ;k)$ & $-ig_vg_sC^0(\mu ,5;k)$ \\ 
$-ig_sg_vC^0(5,\nu ;k)$ & $-g_s^2C^0(5,5;k)$%
\end{tabular}
\right) \text{.}  \tag{19}
\end{equation}
If we substitute $\overline{\Gamma }_\kappa =\{ig_v\gamma _\mu ,g_s\}$ for $%
\Gamma _\kappa $ and use an overhead bar to indicate an expression obtained
in this way, we have $\overline{C}^0(a,b;k)=-\Pi _{ab}^0(k)$. In Eq. (18)
the free meson propagator is given by $\Delta _{ab}^0=\left( 
\begin{tabular}{ll}
$\Delta _v^0\delta _{\mu \nu }$ & $0$ \\ 
$0$ & $\Delta _s^0$
\end{tabular}
\right) $, where $\Delta _\kappa ^0(k)=-i[k^2+m_\kappa ^2-i\epsilon ]^{-1}$
for $\kappa =v$ or $s$, and $m_\kappa $ is determined by $[k^2+m_\kappa
^2+i\Pi _\kappa ^0(k^2;F)|_{M^{*}=M}]_{k^2=-\widehat{m}_\kappa ^2}=0$ with $%
\widehat{m}_\kappa $ denoting the physical mass, and for our present purpose
no medium dependence of meson masses is considered. It will be studied in
more detail elsewhere. If only the summation of ring diagrams is considered,
as illustrated in Fig. 1, RRPA to $C(A,B;kk^{\prime })$ (see Eqs. (2) and
(5a)) may be written as 
\begin{equation}
C(A,B;k,k^{\prime })=C^0(A,B;k,k^{\prime })-\int \frac{d^4q}{(2\pi )^4}%
\widehat{C}^0(A,c;k,q)\Delta _{cd}(q)\widehat{C}^0(B,d;-k^{\prime },-q)\text{%
,}  \tag{20a}
\end{equation}
\begin{equation}
\widehat{C}^0(E,f;k,q)=2\pi \delta (k_0-q_0)\int \frac{d^4p}{(2\pi )^4}%
\mathbf{Tr}[\Gamma _E(\mathbf{k}-\mathbf{q})G(p+q)\overline{\Gamma }_fG(p)]%
\text{,}  \tag{20b}
\end{equation}
where $E=A$ or $B$. Eq. (18) shows the ring summation is achieved by $\Delta
_{cd}$ in Eq. (20). If $\Gamma _E$ is independent of $\mathbf{x}$, from Eq.
(18) one finds easily by iteration that Eq. (20) can be simplified to $%
\overline{C}_{RPA}(a,b;k)=-\Pi _{ab}(k)$ with 
\begin{equation}
\Pi _{ab}(k)=\Pi _{ab}^0(k)+\Pi _{ac}^0(k)\Delta _{cd}^0(k)\Pi _{db}(k)\text{%
.}  \tag{21}
\end{equation}

\section{Numerical results}

We have solved Eq. (21) with $\Pi _{ab}^0$ given by Eq. (19), where the
scalar and vector mesons as well as their mixing are considered. The
parameters are $\overline{g}_s^2=0.5263$, $\overline{g}_v^2=0.6842$ ($%
\overline{g}^2\equiv g^2/16\pi ^2$), whereas $M^{*}=0.73M$, $M=939$, $%
\widehat{m}_s=550$, $\widehat{m}_v=783$ (all in $MeV$), and $k_F=1.3fm^{-1}$
[1].

To illustrate the effects of $C_{FF}^0$ and $\Delta C_m^0$ (see Eqs. (7) and
(12)) as well as the difference in results between the Dirac and D-F
theories, we shall, as an example, discuss $C(4)=Re[C(4,4;k)]=g_v^{-2}Re[\Pi
_{44}(k)]$, which is nonnegative and closely related to the longitudinal
response function. Fig.2 shows the dependence of $C^0(4)$ and $\widetilde{C}%
^0(4)$ on the energy transfer $k_0$. Indeed, in the small $k_0$ region, we
have $C^0(4)=\widetilde{C}^0(4)$, because both $C_{FF}^0(4)$ and $\Delta
C_m^0(4)$ will not be zero only if $k_0>2M^{*}$. Thus, to the lowest order,
the two methods yield the same results in case $k_0<2M^{*}$. In the large $%
k_0$ region since $C_{DD}^0(4)=\widetilde{C}_{DD}^0(4)=0$ (see Eq. (A. 26)
in Ref. [13]) and according to Eq. (14a) $M_1(4,4;k)=0$ for time-like $k_\mu 
$ [11], one gets $\widetilde{C}^0(4)=\widetilde{C}_m^0(4)$, $C_m^0(4)=-%
\widetilde{C}_m^0(4)$ and $C^0(4)=C_m^0(4)+C_{FF}^0(4)$. From Fig. 2 it is
seen that the sign of $\widetilde{C}_m^0(4)$ is correct, while $C_m^0(4)$
has a wrong sign. However, in the Dirac theory the relevant effect is
represented by $C^0(4)$ whose sign is again correct owing to the
contribution of $C_{FF}^0(4)$. It is interesting to find that $\Delta C_m^0$
included in $\widetilde{C}_m^0$ in Eq. (12a) actually means a correction
rather than a defect, though its effect is not very important. For the RRPA
calculation we shall consider three cases: (1) Both $\Gamma _a$ and $\Gamma
_b$ in the Eq. (5) are bare vertices. (2) One of $\Gamma _\kappa $ ($\kappa
=a$ or $b$) is a dressed vertex $\widehat{\Gamma }_\kappa $, which is
phenomenologically taken as 
\begin{equation}
\widehat{\Gamma }_\kappa =\Gamma _\kappa \frac{\Lambda ^2}{\Lambda ^2+|k^2|}%
\equiv \Gamma _\kappa F(k^2,\Lambda )\text{.}  \tag{22}
\end{equation}
(3) Only one $\Gamma _\kappa $ in $C_{FF}^0$ is replaced by $\widehat{\Gamma 
}_\kappa $. For simplicity, the case of dressed vertex at each end of the
loop will not be considered. The detailed formula for $\Pi _{ab}^0$ and the
relevant renormalization procedure are known for all the above three cases
[11-14], thus they will not be written down here. We shall use a subscript $%
\eta =1$ to $3$ to indicate the case which the calculation refers to. In
Fig. 3 the calculated RRPA results for cases (1) and (2) are shown, where we
have chosen $\Lambda =1.1GeV$ for case (2). Since according to Eqs. (18, 19)
and (21) the real as well as the imaginary part of $\Pi _{cd}^0(k)$
contributes, it is seen that there is a significant difference between $%
C_1(4)$ and $\widetilde{C}_1(4)$ even in the small $k_0$ region. Hereafter
we shall restrict our discussion to this region, as the parameters in the
effective Lagrangian are fitted to nuclear ground-state properties. Fig. 3
shows that the difference between $\widetilde{C}_1(4)$ and $\widetilde{C}%
_2(4)$ is noticeably larger than that between $C_1(4)$ and $C_2(4)$. It
shows that the form factor given in Eq. (22) is rather ineffective in the
Dirac case. However, if we assume that the effect of nucleon compositeness
has already been encoded in the effective Lagrangian, we may simply set $%
F(k^2,\Lambda )=1$ i.e. then there would be no need to consider $C_2(4)$ and 
$\widetilde{C}_2(4)$ in order to avoid redundancy. From Eqs. (7) and (19),
it is easily seen that the large difference between $C_1(4)$ and $\widetilde{%
C}_1(4)$ is caused by $C_{FF}^0$ calculated with bare vertices. For this
reason we have further considered case (3), which is represented graphically
in Fig. 4, where the graphs of case (1) are drawn for comparison. We shall
use $C_3(4,\alpha )$ to denote the results calculated with $\widehat{\Gamma }%
_\kappa (\alpha )=\alpha \Gamma _\kappa F(k^2,\Lambda _e)$, where $\alpha $ (%
$=0$ to $1$) is a percentage factor and $\Lambda _e=1.1GeV$. If $\alpha =1$,
we have $C_3(4,1)\simeq C_2(4)\simeq C_1(4)$, i.e. the result of Dirac's
case. On the other hand, if $\alpha =0$, we have $\widehat{\Gamma }_k(\alpha
)=0$, thus $C_3(4,0)=\widetilde{C}_1(4)$ as shown in Fig. 4, i.e. if $%
C_{FF}^0$ or the vacuum contribution is neglected, the Dirac and D-F results
are the same, because the effect of $\Delta C_m^0$ is insignificant. Note
Fig. 4(b) shows that in the case of $k_a=330MeV$ $C_3(4,\alpha )$ varies
with $\alpha $ unusually. It first rises and then falls down finally to $%
C_2(4)\sim C_1(4)$ at $\alpha =1$. Since $C_1(4)\sim C_2(4)$, this suggests
that $\widetilde{C}_2(4,\alpha )$ calculated with $\widehat{\Gamma }_\kappa $
replaced by $\widehat{\Gamma }_\kappa (\alpha )$ may display such a similar
behavior. This is indeed so as shown in Fig. 3(b) where $\widetilde{C}%
_2(4,1)=\widetilde{C}_2(4)$. Clearly, the postulate of filled sea also gives
a greater possibility to realize the experimental data than the ansatz of
empty sea. In fact, if for instance, $C_3(4,\alpha =0.1)$ fits more closely
to the experimental data, it means that the effective Lagrangian has not yet
included the vacuum effect and nucleon compositeness completely and there is
no need to add additional terms and parameters to the Lagrangian, as they
can be taken into account by means of Dirac's method in the simple way. In
the above, we have argued and demonstrated if the effective Lagrangian has
encoded the vacuum contribution and some other higher-order effects (for
instance, nucleon compositeness) in its parameters, we may simply neglect
these contributions in the calculation to avoid redundancy without assuming
that the NE sea is empty and nucleon is structureless.

\section{Summary and conclusions}

To compare the Dirac and D-F methods, we have considered the calculation of
correlation function as a typical example. As shown in section 2, if the
vacuum contribution can be neglected, the Dirac and D-F results are the same
in the small $k_0$ region not only to the lowest order but also in RRPA,
because according to Eq. (16) the effect of $\Delta C_m^0$ is insignificant
in the model studied. Thus, in such cases, the D-F ansatz will yield a good
effective method of calculation, though it is unnecessary, because the Dirac
theory gives the same result and the calculation will be no more complicated
if the vacuum effect and some higher order effects can be neglected. Since
the Dirac theory can identify all the higher order effects unambiguously, to
neglect them we only need to neglect the relevant terms, as displayed in
section 2. Actually, no-sea approximation does not imply that the sea must
be empty. It only means that in the calculation the vacuum contribution can
be neglected, because it has already been encoded in the parameters of the
Lagrangian. As is well known, there is no way to avoid the spontaneous
radiation which will cause MFGS to disappear instantly if the sea is empty.
We have illustrated in sections (2) and (3) why we have emphasized in the
introduction that it is not only important, but also advantageous to follow
Dirac's original postulate of filled sea. Clearly if the effect of $\Delta
C_m^0$ is not small, the D-F and Dirac results may differ, even if the
vacuum contribution can be neglected, thus the difference may then serve as
a discrimination between the two methods. This is a problem under further
study.

This work is supported in part by the National Natural Science Foundation of
China and the Foundation of Chinese Education Ministry.

\newpage\

\newpage\ 

\baselineskip=0.9cm

Figure Captions

Fig. 1: Graphical representation of RRPA for the calculation of the
correlation function $C(A,B;x_1,x_2).$

Fig. 2: Real part of the lowest order approximation to the correlation
function $C(4,4;k)$: $C^0(4)\equiv Re[C^0(4,4;k)]$. In the small $k_0$
region the three curves coincide.

Fig. 3: Curves represent $C(4)\equiv Re[C(4,4;k)]$ in RRPA. The solid curve
denotes $C_1(4)$, the thin one $\widetilde{C}_1(4)$, whereas the heavily
dashed one is $C_2(4)$, and the dashed one $\widetilde{C}_2(4,\alpha )$
where $\widetilde{C}_2(4,1)=\widetilde{C}_2(4)$. (both with $\Lambda =1.1GeV$%
). (a) momentum transfer $k_a\equiv |\mathbf{k}|=560MeV$, (b) $k_a=330MeV$.

Fig. 4: Effects of vacuum contribution and nucleon compositeness in RRPA.
The solid and thin curve again reprint $C_1(4)$ and $\widetilde{C}_1(4)$,
respectively. The dashed curves attached with the value of $\alpha $ ($%
\alpha =0.3$, $0.6$ and 1) denote the respective $C_3(4)$. (a) $k_a=560MeV$,
(b) $k_a=330MeV$.

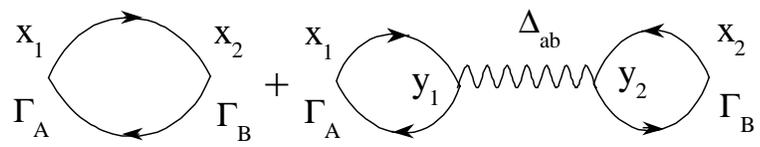
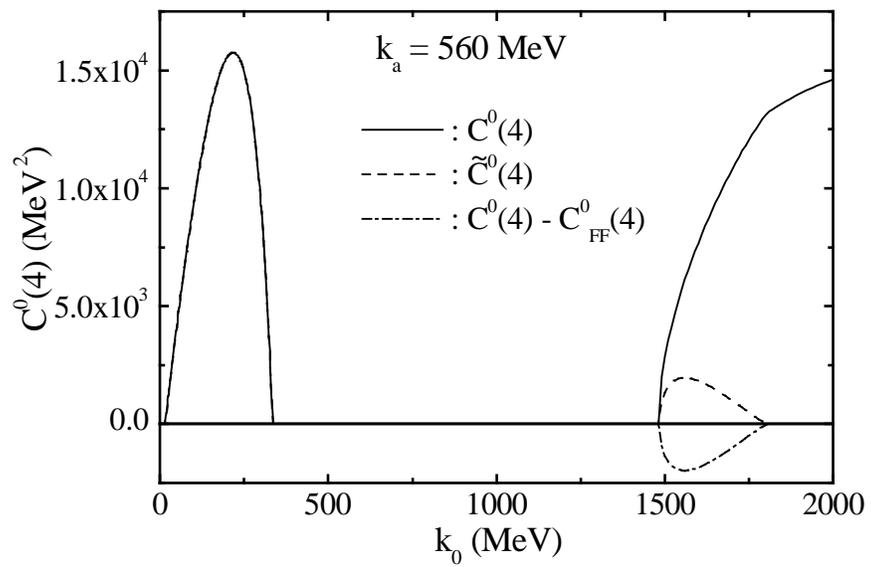
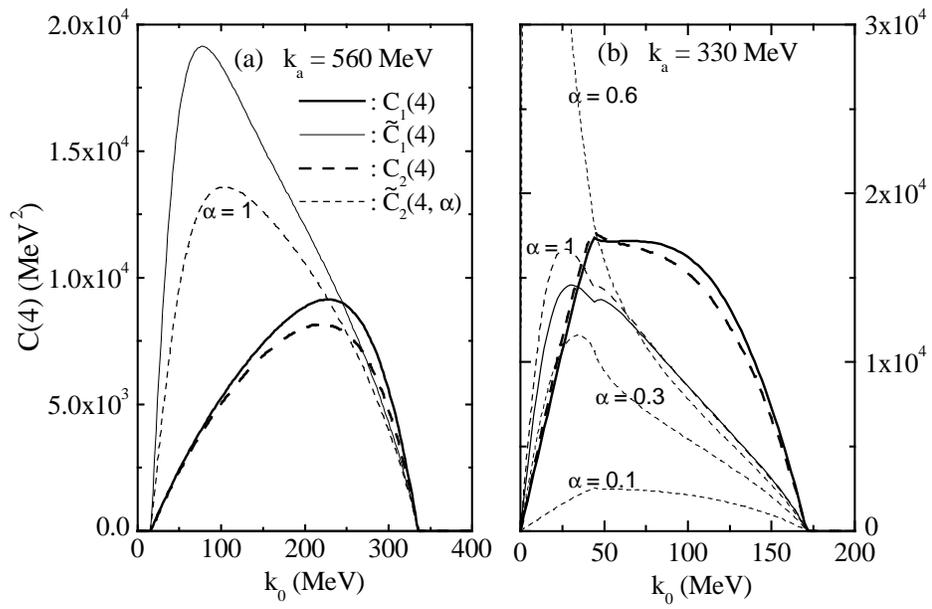
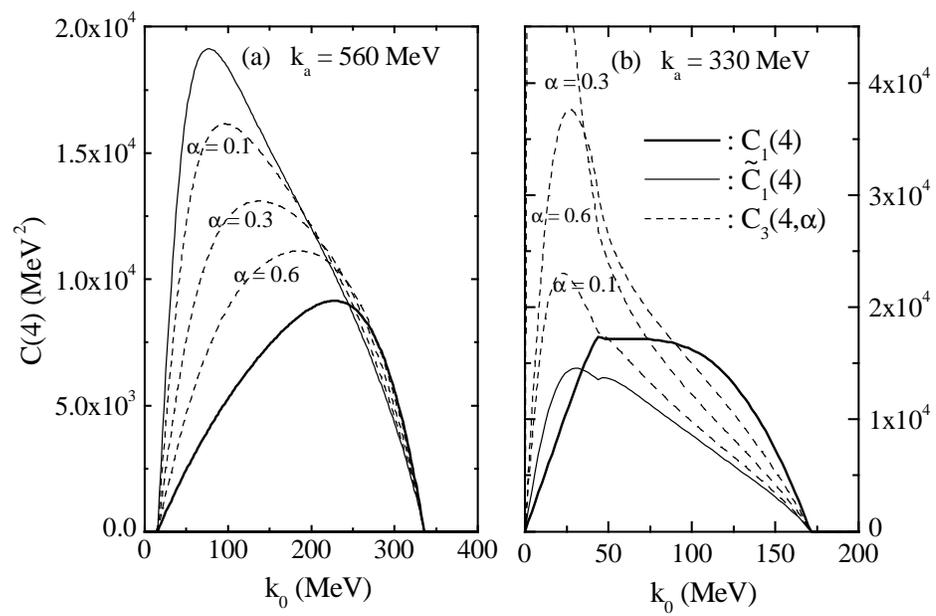
\end{document}